\documentclass[10pt,conference]{IEEEtran}

\usepackage{cite}
\usepackage{graphicx}
\RequirePackage{epsfig}
\usepackage{xspace}

\usepackage{amsmath}
\usepackage{amssymb}
\usepackage{algorithm}
\usepackage{algorithmic}
\usepackage{amsfonts}
\usepackage{amscd}

\renewcommand{\algorithmiccomment}[1]{#1}

\begin{document}

\title{A Two Stage Selective Averaging LDPC Decoding}
  
\author{\authorblockN{Dinesh Kumar. A , Ambedkar Dukkipati} 
\authorblockA{Department of Computer Science and Automation\\
Indian Institute of Science, India\\
Email: \{dinesh.a, ambedkar\}@csa.iisc.ernet.in}
}

% make the title area
\maketitle

\begin{abstract}
Low density parity-check (LDPC) codes are a class of linear block codes that are decoded by running belief 
propagation (BP) algorithm or log-likelihood ratio belief propagation (LLR-BP) over the factor graph of the code. One of the disadvantages of LDPC codes is the onset of an error floor at
high values of signal to noise ratio caused by trapping sets.  In this paper, we propose a two stage decoder to deal with different types of trapping sets. 
Oscillating trapping sets are taken care by the first stage of the dcecoder and the elementary trapping sets are handled by the second stage of the decoder. 
Simulation results on regular PEG (504,252,3,6) code shows that the proposed two stage decoder performs significantly better than the standard decoder.
\end{abstract}

\section{Introduction}
       \noindent
Low Density Parity-Check (LDPC) codes were proposed by R. Gallager in 1963 \cite{galla_ldpc} and was rediscovered by Mackay \cite{mackay_ldpc} in late
1990s. Since they are based on sparse parity-check matrix, iterative decoders becomes an attractive option. Due to the 
near-capacity performance and the low complexity of the decoder they have become an intensive research topic among the
coding theory community.

A LDPC code can be described by a parity-check matrix $ \mathbf{H}$ that is represented using  a special type of graph called factor graph. A factor graph is a 
bipartite graph with two different types of nodes, variable nodes and check nodes corresponding to the columns and rows of the 
parity-check matrix $\mathbf{H}$. H$(i,j) = 1$ corresponds to an edge between the variable node $v_j$ and check node $c_i$. 

Unlike the maximum a-poteriori probability (MAP) decoding algorithm that seeks for the global optimization over the whole code word space, belief propagation (BP) algorithm seeks for a local optimization by using only the information that is 
flowing through the variable nodes, without the knowledge of the global state. BP algorithm achieves this by passing messages iteratively between 
the variable nodes and check nodes. The message $m_{v_ic_j}^{(l)}$ at iteration $l$ 
from the variable node $ v_i $ to the check node $c_j $ can be interpreted as the probability that the variable node $v_i$ is in state $ x $ in the absence of 
the check node $c_j$. Similarly, the message $m_{c_jv_i}^{(l)}$ at iteration $l$ from the check node $c_j$ to variable node $v_i$ is the probability
that the check node $c_j$ is satisfied in the absence of variable node $v_i$.

The main problem with LDPC codes is that  these codes tend to exhibit a sudden saturation for sufficiently high signal to noise ratio (SNR).
Trapping sets \cite{richard_ldpc} are considered the primary reason for this behavior of LDPC codes. 
In this paper, we propose a two stage decoder that in the first stage averages the messages from the selected variable nodes over two iterations and the nodes are selected if the 
belief in the node is decreasing below a certain threshold or if the belief is increasing rapidly. The decoded string is checked if it has converged to a valid code word in
each iteration. If the decoded string has not converged to a valid code word even after maximum number of iterations and the number of 
check nodes that are not satisfied is below a certain threshold, we proceed to the second stage of the decoder where we flip certain bits connected to the unsatisfied check nodes. 
Then the first stage of the decoding process is repeated for the processed string. 
The selective averaging of the messages in the first stage slows down the information flow from the nodes affected by the trapping sets and the remaining variable nodes are not affected. These reliable nodes 
grow in belief and helps in solving the oscillating trapping sets. The elementary trapping sets are an unstable equilibrium, flipping some of the bits in trapping set will break the trapping set. This is the 
intuition behind the second stage of the proposed decoder.

The paper is organized as follows. \S~ \ref{Section:Decoding Algorithm} explains the basic decoding algorithm. 
 \S~ \ref{Section:Trapping Sets} contains a description of trapping sets and how they affect the performance of the 
decoding BP algorithm. \S~ \ref{Section:Related Work} explains the various methods proposed in literature for improving the decoder performance.
\S~ \ref{Section:Proposed Algorithm} proposes our two stage decoder and analyses how they improve the performance from the 
trapping set point of view. \S~ \ref{Section:Simulation Results} gives the simulation results to prove 
that the proposed algorithm does better than averaging decoder and standard decoder. We provide the concluding remarks in  \S~ \ref{Section:Conclusion}.

\section{Background}
\label{Section:Background}

\subsection{Decoding Algorithm}
  \label{Section:Decoding Algorithm}
       \noindent

The parity check matrix of the LDPC codes are represented by a bipartite factor graph composed of N variable nodes $v_{j}$, for $ j \in \{1,\ldots,N\}$, that 
represent the message bits in the codeword and M check nodes $c_{i}$, for $i \in \{1,\ldots,M \}$, that represents the 
parity-check equations. The channel considered here is the additive white Gaussian noise (AWGN) channel. There are two types of 
decoding algorithms . The standard belief propagation (BP) algorithm which uses product of probabilities as messages and the other 
being a modified version of BP algorithm which uses loglikelihood ratio (LLR) as the messages between the nodes.In general the log
likelihood ratio belief propagation (LLR-BP) algorithm is prefered for computational accuracy. The sign of LLR indicates the most likely value of the bit (0 or 1) and 
the absolute value of the LLR gives the reliability of the message.

Let $m_{v_{i}c_{j}}^{(l)}$ be the message passed from a  variable node $v_{i}$ to a check node $c_{j}$ at the $ l ^{th}$ iteration of the algorithm. 
Similarly, one can define $m_{c_{j}v_{i}}^{(l)}$. The initial values of the message $m_{v_{i}c_{j}}^{(0)}$ is equal to the channel information 
LLR $C_{v_{i}}$ of the variable node $v_{i}$ and it is independent of $c_{j}$. We have

\[ 
  m_{v_{i}c_{j}}^{(l)} = C_{v_{i}} + \sum_{c_{a} \in N(v_{i}) \backslash c_j} m_{c_{a}v_{i}}^{(l-1)}\mbox{\space,}
 \]
and
\[
  m_{c_iv_j}^{(l)} = 2 \times \operatorname{atanh} \left( \prod_{v_b \in N(c_i)\backslash v_j} \tanh \left( \frac{m^{(l-1)}_{v_bc_i}}{2} \right) \right)\mbox{\space,}
\]
where $N(v_i) \backslash c_j$ denotes the neighbors of $v_i$ excluding $ c_i$ and $ N(c_i)\backslash v_j$ denotes the neighbors
of $c_i$ excluding $v_j$.

The algorithm follows the flooding schedule in which an iteration consists of simultaneous update of all the 
messages $m_{v_ic_j}$, followed by the simultaneous update of all the messages $m_{c_jv_i}$. The decoding algorithm stops if the
decoded bits satisfy all the parity-check equation or maximum number of iterations is reached.

\subsection{Trapping sets}
  \label{Section:Trapping Sets}
       \noindent

The notion of trapping set was described in the context of error-floor analysis in \cite{MacKayPostol2002}. This concept was
further developed in the seminal paper by Richardson \cite{richard_ldpc}, in which trapping set configurations were shown to exhibit a strong influence on the point 
of onset as well as on the slope of the error-floor curve of LDPC codes.
\newline

Trapping set is defined as follows. An $(x,y)$ trapping set $\tau$ is a configuration of $x$ variable nodes, for which the induced subgraph contains $y \ge 0$ 
odd degree check nodes.\newline

\begin{figure}[!h]

\begin{center}
%left,bottom,right,top
\includegraphics[scale = 0.5]{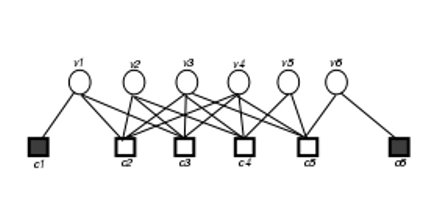}
\end{center}
\caption{(6,2) Trapping set} \label{fig:trapps}
\end{figure}

An example of a trapping set can be seen in Figure \ref{fig:trapps}. In the above figure we can see six variable nodes and two unsatisfied 
check nodes. A trapping set always has some unsatisfied check nodes. This is how we differentiate the error due to trapping set and the
error that occurs due to minimum distance of the codes.

There are three types of error events in the error floor region \cite{jour_ldpc}. They are
\begin{enumerate}
 \item unstable error events, which dynamically change from iteration to iteration and for which $x$ and $y$ are typically large,
 \item stable or elementary trapping sets, for which $x$ and $y$ are small and are the main cause of the error floor and 
 \item oscillating trapping sets, which periodically vary with the number of decoder iterations and which are some times
  subsets of stable trapping sets.
\end{enumerate}

As explained in \cite{algo_analysis_ldpc}, the onset of of error-floors in LDPC codes can be attributed to the following phenomena. In the initial stages of the
belief propagation, due to presence of low-probability noise samples, variable nodes internal to one particular trapping
set (\textit{initial trapping set}) experience a large increase in belief towards the wrong bit value. This information
gets propagated to other variable nodes in the trapping set, some of which already have very 
low belief in their bit value. After this initial increase in belief, external variables usually start adjusting the incorrect 
estimates towards the correct bit values. By this time the variable nodes in the trapping sets that are affected by the 
inital information surge are significantly biased towards the incorrect bit values. Since there are only a few check nodes capable of 
detecting errors within the trapping sets, this incorrect bit estimates remain unchanged until the end of the decoding process.

The trapping sets are union of several cycles. One of the most prevalent types of error in the waterfall region is the oscillating trapping sets 
which causes the LLR values of a node to oscillate. This could be caused by two or more cycles passing through a node. When the 
messages looping in the cycles supporting different bit values arrive at the variable node at different times,
the LLR values seems to oscillate.

\subsection{Related Work}
  \label{Section:Related Work}
       \noindent
A lot of research has gone into designing decoders to mitigate the  errors caused by the trapping sets. The decoders are
designed in such a way that the computational complexity of the modified decoder is not high compared to the standard decoder.

In \cite{algo_analysis_ldpc}, the problems caused by the trapping sets and why the decoder fails in overcoming these problems are well
studied. Landner et al. \cite{algo_analysis_ldpc}, \cite{Laendner_arvix} propose an averaging decoder that averages the LLR value of the node over several iterations. 
Averaging prevents the erroneous information from being trapped in the code graph by slowing down the convergence speed of the nodes. This method though computationally less complex,
it slows down the convergence of the reliable nodes. This affects the performance of the decoder in the waterfall region as oscillating 
trapping sets are prevalent in this region than the elementary trapping sets. 

In \cite{improve_bp}, the BP algorithm is well studied in the point of view of the Bethe energy and how it fails in the 
presence of cycles. They propose a BP algorithm with a tunable parameter $ \Delta$ and a modification to the outgoing message from the 
variable node. On adjusting the parameter $ \Delta$ at different SNR points, they are able to achieve better performance  than the standard BP 
decoder. This modification also follows the same principle as the averaging decoder by slowing down the information flow to prevent 
the trapping of the erroneous information.

The work put forward in \cite{oscillation_ldpc} concentrates on short and middle length LDPC codes as they are largely 
affected by cycles in the graph. These cycles causes the oscillation of the LLR values in the nodes. The messages from the previous iteration is 
added to the message in the current iteration if the sign of the message changes. This modification helps in damping the oscillation, but their method does not handle the errors caused by elementary trapping sets.

A two stage backtracking decoder was proposed in  \cite{two_stage} \cite{ldpc_backtrack}. The fact that an unsatisfied check node is connected to odd number of variable nodes in trapping set is 
used to construct a matching set $\Psi$. Then each variable node that belong to the matching set $\Psi$ is flipped seperately and the first stage is run. This process is repeated till the error is solved or 
all the nodes in the matching set have been exhausted. 

\section{Proposed Algorithm}
  \label{Section:Proposed Algorithm}
       \noindent

The proposed decoding algorithm consists of two stages. The first stage is selective averaging decoding in which we concentrate on 
the oscillating trapping sets. If the decoder does not converge after a fixed number of iterations and 
the number of unsatisfied check nodes is below a certain threshold, we assume that it is due to elementary trapping sets and identify the 
variable nodes that are connected to the unsatisfied check node and multiply the channel information LLR $C_{v_i}$
with a constant and repeat the first stage of the decoder.

Our aim in the first stage is to prevent the error due to oscillating trapping sets and initial trapping set. One has to  take care that the
algorithm does not affect the convergence of reliable nodes. Faster convergence of the reliable nodes seems to help 
the convergence of the oscillating nodes as the messages from the converged reliable nodes are very strong compared to the belief in the oscillating nodes.
As explained in \S~ \ref{Section:Trapping Sets}, the nodes affected by the initial trapping set
have a rapid increase in reliability value on the wrong bit and the errors caused by this can be prevented by slowing down the wrong information from flowing out of these 
nodes. 

The selective averaging algorithm modifies the outgoing messages $m^{(l)}_{v_ic_j}$ from the variable nodes as follows.

\[
 m'^{(l)}_{v_ic_j} = 
\begin{cases}
 \frac{m^{(l)}_{v_ic_j} + m^{(l-1)}_{v_ic_j}}{2}, & \mbox{if}\; Selected(v_i) > 0 \\
 m^{(l)}_{v_ic_j}, & \mbox{if}\; Selected(v_i) = 0\\
\end{cases}
\]

The messages from the check nodes $m^{(l)}_{c_jv_i}$are not modified.

\begin{algorithm}
\caption{Node Selection} \label{alg:node_selection}
\begin{algorithmic}
\FOR { $i \in {1,2, \ldots , N} $ }
\STATE $B^{(l)}_{(v_i)} \gets | L^{(l)}_{(v_i)} |$
\STATE $B^{(l-1)}_{(v_i)} \gets | L^{(l-1)}_{(v_i)} |$
\IF {$ Selected(v_i) != 2 $}
  \STATE $Selected(v_i) \gets 0 $
\ENDIF
\IF {$B^{(l)}_{(v_i)} < B^{(l-1)}_{(v_i)} \; \&\& \; (B^{(l-1)}_{(v_i)} - B^{(l)}_{(v_i)}) > {\beta}$} 
        \STATE $Selected(v_i) \gets 1$
\ELSIF {$B^{(l)}_{(v_i)} > B^{(l-1)}_{(v_i)}$}
    	\IF{$ (B^{(l)}_{(v_i)} - B^{(l-1)}_{(v_i)}) > {\nu} $}
	    \STATE $Selected(v_i) \gets 1$
	\ENDIF
\ENDIF 

\ENDFOR
\end{algorithmic}
\end{algorithm}

Algorithm \ref{alg:node_selection} proposes the method for selection of nodes. 
$L^{(l)}_{(v_i)}$ represents the log likelihood ratio of the node $v_{i}$ at iteration $l$ and $L^{(l-1)}_{(v_i)}$ 
represents the log likelihood ratio of the node $v_{i}$ at iteration $l-1$. We assign  $| L^{(l)}_{(v_i)} |$ to 
$B^{(l)}_{(v_i)}$ and $| L^{(l-1)}_{(v_i)} |$ to $B^{(l-1)}_{(v_i)}$. We then check if the belief in the node $B_{(v_i)}$ is 
decreasing or increasing with the iteration.  If the belief is decreasing below a constant $\beta$, we assume this as an oscillating 
node and the node is selected. If the belief is increasing rapidly at a rate above the constant $\nu$, we consider this behaviour as 
an initial trapping set and the node is selected for averaging of the messages in the next iteration. This condition also 
helps prevent the oscillating nodes. All the nodes are unselected at the beginning of the node selection procedure at each iteration except the 
nodes which are modified in the second stage of the algorithm. These nodes are assigned  $Selected(v_i) = 2$ and the messages from these 
nodes are always averaged. A more detailed explanation for this special treatment of the modified nodes will follow when we 
explain the second stage of the algorithm.

The first stage of the decoder is not designed for dealing with the elementary trapping set which is the most prevalent type of trapping set in the 
error floor region. We will use a similar algorithm to the one proposed in \cite{two_stage}. The algorithm proposed in 
\cite{two_stage} \cite{ldpc_backtrack}, uses a backtracking approach and forming a set of matching sets of vertices which might belong to the trapping set. The initial LLR value 
of the variable nodes that belong to one of the members of the set of matching set is flipped and the decoding process is repeated. If the trapping set error is not solved, then 
the decoder backtracks and repeats the process for the next member of the set of matching set. The second stage of our 
decoder also follows a similar approach but we avoid backtracking as it increases the complexity of the decoder.

\begin{algorithm}
\caption{Second Stage of the decoder} \label{alg:second_stage}
\begin{algorithmic}[1]

\STATE First stage decoder unsuccessful.
\STATE Find the set $C$ of unsatisfied check nodes.
\IF {$ |C| < CN_{threshold} $}

  \STATE $Vhat_{old} \gets $output of first stage of the decoder. 
  \STATE Find the set of variable nodes $V_{un}$ where $V_{un} = \bigcup \limits_{c_j \in C} N(c_j)$.
 
  \algorithmiccomment $N(c_j)$ is the set of all variable nodes which are connected to the check node $c_j$. 

  \STATE Find the subset of variable nodes $V_{nc} \subseteq V_{un}$ such that no two variable nodes have common check nodes other than the unsatisfied check nodes.
  \FOR {$v_k \in V_{nc}$}
    \STATE $C'_{(v_k)} =  -C_{(v_k)} * \eta $

    \algorithmiccomment $C_{(v_k)}$ is the channel information LLR of the variable $v_k$
    \STATE $Selected(v_k) \gets 2$
  
  \ENDFOR
\ENDIF
\STATE re-decode with the the new channel information LLR $C'_{(v_k)}$.

\IF{re-decode is successful}
  \STATE stop and exit
\ELSE
 \STATE $Vhat_{new} \gets $output of re-decode. 
  \STATE Find the set $C'$ of unsatisfied check nodes. 
  \IF{$|C| < |C'|$}
   \STATE  output $Vhat_{old}$.
  \ELSE
    \STATE output $Vhat_{new}$.
  \ENDIF
\ENDIF

\end{algorithmic}
\end{algorithm}

Algorithm \ref{alg:second_stage} lists the second stage of the proposed decoder. If the first stage of the decoder has not converged even
after the maximum number of iterations and the number of unsatisfied check nodes is below $CN_{threshold}$ which is a very small 
number (Trapping set errors are low weight errors in error floor region), we conclude that the error occured is due to an
elementary trapping set. As the trapping set is an unstable equilibrium condition, if we flip even one or two variable nodes 
that belong to the check nodes, it is enough to break the trapping set. But we do not know which variables nodes belong to the 
trapping sets. The only information that is available to us is that the unsatisfied check nodes are connected to odd number of variable nodes 
(one node in the worst case) in the trapping set as shown in Figure \ref{fig:trapps}. We find the set of variables nodes  $V_{un}$ that is a 
union of all variable nodes connected to the unsatisfied check nodes. Find a subset $V_{nc} \subseteq V_{un}$ such that no two variable nodes 
in the set $V_{nc}$ has a common check node other than the unsatisfied check nodes. We will multiply the initial LLR value $C_{v_k}$ with -1 and 
$\eta$. This flips the bit value of the node and the decoding process is repeated. But some of the correctly decoded nodes which do not belong 
the trapping set might also be flipped. Since the variable nodes do not have any common check nodes, the wrongly flipped nodes are easily corrected by other 
reliable nodes in the decoding process. If the re-decoding is not successful, then we report the string with the least number of unsatisfied check nodes.
To be cautious we assign $Selected(v_k) = 2$ so that they remain always selected and 
the messages from these nodes are always averaged. This helps in preventing the wrong information from spreading intially before the 
reliable nodes converge. The nodes that we have flipped in this stage that belong to trapping set will help in breaking the trapping set.

\section{Simulation Results}
  \label{Section:Simulation Results}
       \noindent
We ran simulations of the proposed decoder for (504,252,3,6) regular PEG code at different SNR points. We compare the results of the proposed decoder to the 
averaging decoder \cite{algo_analysis_ldpc} and standard decoder. Without loss of generality, the codewords send were all-zero.

\begin{figure}[!h]
\begin{center}
%left,bottom,right,top
\includegraphics[scale = 0.5]{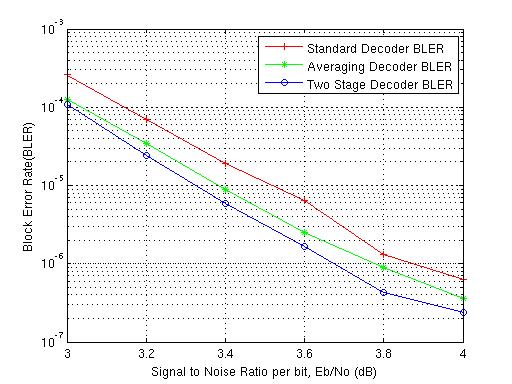}
\end{center}
\caption{Comparison of BLER (504,252,3,6) regular PEG code} 
\label{fig:final_bler}
\end{figure}

\begin{figure}[!h]
\begin{center}
%left,bottom,right,top
\includegraphics[scale = 0.5]{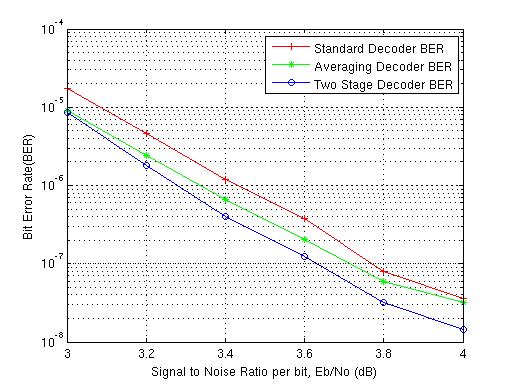}
\end{center}
\caption{Comparison of BER (504,252,3,6) regular PEG code} 
\label{fig:final_ber}
\end{figure}

In Figure \ref{fig:final_bler}, \ref{fig:final_ber} we can see that the proposed two stage decoder performs better than the  other two decoders in both bit error rate (BER) and (BLER). This performance is 
expected because the proposed two stage decoder solves the trapping sets better than the averaging decoder \cite{algo_analysis_ldpc} and standard decoder.

\begin{figure}[!h]
\begin{center}
%left,bottom,right,top
\includegraphics[scale = 0.5]{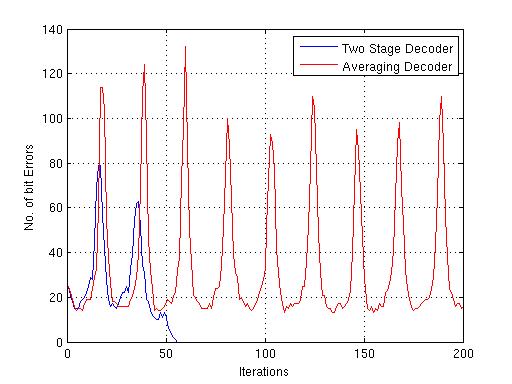}
\end{center}
\caption{Comparison of the error performance of two stage decoder and averaging decoder.} 
\label{fig:two_compare}
\end{figure}

In Figure \ref{fig:two_compare} we can see the number of bit errors in  the decoder after processing in 
the second stage of the decoder and the number of bit errors in the averaging decoder. We can see that the second stage of the proposed decoder 
helps in solving the error due to the elementary trapping sets.

We can improve the second stage of the decoder by decreasing the number of bits flipped that do not belong to the trapping sets. We are also working on error-floor
prediction methods which will be useful for high SNR value where simulation is not a possibility.

\section{Conclusion}
  \label{Section:Conclusion}
       \noindent
In this paper, we proposed a two stage decoding algorithm for LDPC codes that reduces the error performance of the code especially in the error 
floor region. In the first stage we handle the oscillating trapping sets by selecting the nodes based on rate of increase or decrease of the belief. 
The messages from the selected nodes are averaged with message from the previous iteration to slow down the flow of erroneous 
information in the system. The reliable nodes are allowed to converge faster and they help in solving the oscillating trapping sets.
The first stage decoder does not solve the errors due to elementary trapping set. To address this we added a second stage 
to the decoder that flips bits that are connected to the unsatisfied check nodes. This helps in breaking the trapping sets.

\bibliographystyle{IEEE}

\bibliography{pub}
\end{document}